\documentclass[twocolumn,preprintnumbers, amssymb,amsmath,aps,floatfix,prl,nofootinbib,superscriptaddress,showpacs]{revtex4-1}
\usepackage{mathrsfs}
\usepackage{amsfonts}
\usepackage{amsmath}
\usepackage{array}
\usepackage{verbatim}
\usepackage{epsfig}
\usepackage{graphicx}
\usepackage{hyperref}

\usepackage{xcolor}

\begin{document}
\title{Lepton-jet Correlations in Deep Inelastic Scattering at the Electron-Ion Collider}

\author{Xiaohui Liu}
\affiliation{Center of Advanced Quantum Studies, Department of Physics,
Beijing Normal University, Beijing 100875, China}

\author{Felix Ringer}
\affiliation{Nuclear Science Division, Lawrence Berkeley National
Laboratory, Berkeley, CA 94720, USA}

\author{Werner Vogelsang}
\affiliation{Institute for Theoretical Physics,
                Universit\"{a}t T\"{u}bingen,
                Auf der Morgenstelle 14,
                D-72076 T\"{u}bingen, Germany}
                
\author{Feng Yuan}
\affiliation{Nuclear Science Division, Lawrence Berkeley National
Laboratory, Berkeley, CA 94720, USA}

\begin{abstract}
We propose the lepton-jet correlation in deep inelastic scattering  as a unique tool for nucleon/nucleus tomography at the electron-ion collider. The azimuthal angular correlation between the final state lepton and jet depends on the transverse momentum dependent quark distributions. We take the example of single transverse spin asymmetries to show the sensitivity to the quark Sivers function. When the correlation is studied in lepton-nucleus collisions, transverse momentum broadening effects can be used to explore cold nuclear matter effects. These features make lepton-jet correlations an important new hard probe at the EIC.
\end{abstract}

\maketitle

{\it 1. Introduction.}
A high energy and high luminosity polarized electron-ion collider (EIC) is regarded as the next generation QCD machine where the partonic structure of nucleons and nuclei will be explored in great detail~\cite{Boer:2011fh,AbelleiraFernandez:2012cc,Accardi:2012qut}. In this paper, we propose lepton-jet correlations as a new class of observables in deep inelastic scattering (DIS). They will provide unique probes of various interesting physics aspects and 
thus set a new direction for the EIC science program.

\begin{figure}[t]
\begin{center}
\includegraphics[width=6cm]{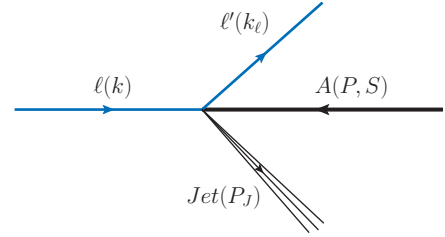}
\end{center}
\caption[*]{Lepton-jet correlation for the tomography of the nucleon/nucleus at the EIC.}
\label{fac0}
\end{figure}

We focus on large transverse momentum lepton-jet production in the center of mass (CM) frame of the incoming lepton and nucleon, see Fig.~\ref{fac0},
\begin{equation}\label{eq:process}
\ell (k)+A(P)\to \ell'(k_\ell)+\text{Jet} (P_J)+X \, ,
\end{equation}
where the lepton and nucleon carry momenta $k$ and $P$ and we denote the momenta of the outgoing lepton and jet by $k_\ell$ and $P_J$, respectively. We label the rapidities of the final state lepton and jet as $y_\ell$ and $y_J$ and their transverse momenta as $k_{\ell\perp}$ and $P_{J\perp}$. All of these kinematic variables are defined in the CM frame of the incoming lepton and nucleon. This is very different from the jet measurements in previous DIS experiments such as those carried out at HERA~\cite{Abramowicz:2017ful,Abramowicz:2012jz,Abramowicz:2010ke}, where the cross sections were measured in the CM frame of the virtual photon and nucleon. Similar studies at hadron colliders have been carried out previously for the correlation of dijets~\cite{Abazov:2004hm,Khachatryan:2011zj,daCosta:2011ni,Adamczyk:2013jei,Adamczyk:2017yhe}. At the EIC, the lepton-jet correlation depends on much simpler kinematics, and at the same time utilizes the observed jet as an important probe of the nucleon/nucleus, as will be demonstrated in this paper.

At leading order, the final state lepton and jet are produced back-to-back in the plane transverse to the beam direction. The intrinsic transverse momentum of the quark and higher order gluon radiation will induce an imbalance between the final state particles. In the correlation limit where the imbalance transverse momentum $q_\perp=|\vec{k}_{\ell\perp}+\vec{P}_{J\perp}|$ is much smaller than the lepton transverse momentum, we can factorize the differential cross section into the transverse momentum dependent (TMD) quark distribution~\cite{Collins:1981uk,Collins:1981uw,Collins:1984kg,Ji:2004wu,Collins:2011zzd,GarciaEchevarria:2011rb} and the soft factor associated with the final state jet,
\begin{eqnarray}
\frac{d^5\sigma(\ell p\to \ell' J)}{dy_\ell d^2k_{\ell\perp} d^2q_\perp}=
\sigma_0 \int d^2k_\perp d^2\lambda_\perp xf_q(x,k_\perp,\zeta_c,\mu_F)\nonumber\\
\times H_{\mathrm{TMD}}(Q,\mu_F)S_J(\lambda_\perp,\mu_F)\,
\delta^{(2)}(q_\perp-k_\perp-\lambda_\perp) \ .
\end{eqnarray}
Here $f_q$ represents the TMD quark distribution, $S_J$ is the soft function associated with the final state jet which implicitly also depends on the jet radius $R$, and $H_{\mathrm{TMD}}$ is the hard factor. In the above, $\sigma_0$ is the leading order cross section, $\mu_F$ denotes the factorization scale, and $\zeta_c$ is the rapidity cutoff parameter needed in order to define the TMD quark distribution~\cite{Collins:2011zzd}. We emphasize that the process introduced here is complementary to semi-inclusive hadron production in DIS processes (SIDIS)~\cite{Mulders:1995dh,Boer:1997nt,Bacchetta:2006tn}, where both the TMD quark distribution and fragmentation functions are involved. The uniqueness of the process proposed  in (\ref{eq:process}) is that the lepton-jet correlation is defined in the lab frame, so that data can be compared directly to similar dijet measurements at RHIC and the LHC. This comparison will help in particular to investigate the difference between the hot and cold dense nuclear matter when an energetic jet traverses the QCD medium. 

The remainder of this paper is organized as follows. In Sec. 2, we will derive the TMD factorization for the lepton-jet correlation. In Sec. 3 and 4, we illustrate the powerful reach of this process by considering the Sivers asymmetry in $ep$ collisions and $P_T$-broadening effects in $eA$ collisions. Finally, we summarize our paper in Sec. 5.

{\it 2. TMD Factorization.}
It is convenient to write the factorization in Fourier transform $b_\perp$-space,
\begin{equation}
\frac{d\sigma}{dy_\ell d^2 k_{\ell \perp} d^2q_{\perp}}=\sigma_0\int\frac{d^2b_\perp}{(2\pi)^2}{\mathrm{e}}^{iq_\perp\cdot b_\perp} \widetilde{W}_q(x,b_\perp) \ ,
\end{equation}
where $\sigma_0=\frac{\alpha_e^2e_q^2}{\hat s Q^2}\frac{2(\hat s^2+\hat u^2)}{Q^4}$ with $\hat s$, $\hat t$ and $\hat u$ the usual Mandelstam variables for the partonic subprocess $\ell+q\to \ell '+q$. We have $Q^2=-\hat t =    k_{\ell\perp}\, \sqrt{S_{ep}} \, {\mathrm{e}}^{-y_\ell}$ and $\hat{u} = - x k_{\ell\perp}\, \sqrt{S_{ep}}\,  {\mathrm{e}}^{y_\ell}$,
with $S_{ep}$ the CM energy squared of the incoming lepton and nucleon. Within TMD factorization, we can write $\widetilde{W}_q$ as
\begin{equation}\label{eq:TMDfactorization}
    \widetilde{W}_q
    =xf_{q}(x,b_\perp,\zeta_c,\mu_F)\, S_J(b_\perp,\mu_F)\,  H_{\mathrm{TMD}}(Q,\mu_F)\ .
\end{equation}
where the hard factor $H_{\mathrm{TMD}}(Q,\mu_F)$ depends on the hard momentum $Q$ and $\mu_F$ is the factorization scale. 

It is known that TMD factorization in dijet production in hadronic collisions is more complicated~\cite{Boer:2003tx,Qiu:2007ey,Collins:2007nk,Rogers:2010dm,Bacchetta:2005rm,Vogelsang:2007jk,Bomhof:2007su,Catani:2011st,Mitov:2012gt,Schwartz:2017nmr,Schwartz:2018obd}. However, in our case there are only final state interaction effects, and the TMD quark distribution can be defined in the same way as for SIDIS:
\begin{eqnarray}
f_q^{\mathrm{unsub.}}(x,k_\perp)=\frac{1}{2}\int
        \frac{d\xi^-d^2\xi_\perp}{(2\pi)^3}{\mathrm{e}}^{-ix\xi^-P^++i\vec{\xi}_\perp\cdot
        \vec{k}_\perp}  \nonumber\\
    \times\left\langle
PS\left|\overline\psi(\xi){\cal L}_{n}^\dagger(\xi)\gamma^+{\cal L}_{n}(0)
        \psi(0)\right|PS\right\rangle\ ,\label{tmdun}
\end{eqnarray}
with the future-pointing gauge link $ {\cal L}_{n}(\xi) \equiv \exp\left(-ig\int^{\infty}_0 d\lambda\, v\cdot A(\lambda n +\xi)\right)$. The above definition contains a light-cone singularity, whose regularization and subtraction defines the scheme. However, the final results do not depend on the scheme when the QCD evolution and resummation are performed~\cite{Catani:2013tia,Prokudin:2015ysa}. We apply the ``Collins-11'' scheme for the TMDs with the following soft factor subtraction in $b_\perp$-space~\cite{Collins:2011zzd},
\begin{equation}
\tilde{f}_{q}(x,b_\perp,\zeta_c,\mu_F)=\tilde{f}_q^{\mathrm{unsub.}}(x,b_\perp)\sqrt{\frac{S_{\bar
n,v}(b_\perp)}{S_{n,\bar n}(b_\perp)S_{n,v}(b_\perp)}} \ . \label{jcc}
\end{equation}
Here $\zeta_c^2=x^2(2v\cdot P)^2/v^2=2(xP^+)^2{\mathrm{e}}^{-2y_n}$ where $y_n$ is the rapidity cutoff in the Collins-11 scheme. The second factor represents the soft factor subtraction where $n$ and $\bar n$ are light-front vectors $n=(1^-,0^+,0_\perp)$, $\bar n=(0^-,1^+,0_\perp)$, whereas $v$ is an off-light-front vector $v=(v^-,v^+,0_\perp)$ with $v^-\gg v^+$. The light-cone singularity in the un-subtracted TMD is canceled by the soft factor as in Eq.~(\ref{jcc}) with $S_{v_1,v_2}$ defined as
\begin{equation}
S_{v_1,v_2}(b_\perp)={\langle 0|{\cal L}_{v_2}^\dagger(b_\perp) {\cal
L}_{v_1}^\dagger(b_\perp){\cal L}_{v_1}(0){\cal
L}_{v_2}(0)  |0\rangle   }\, . \label{softg}
\end{equation}
In the process (\ref{eq:process}), soft gluon radiation associated with the jet will also contribute to the imbalance $q_\perp$. This contribution depends on the jet size $R$~\cite{Banfi:2008qs,Mueller:2013wwa,Sun:2014gfa,Sun:2015doa}, which we compute using the narrow jet approximation~\cite{Mukherjee:2012uz}. We also introduce a subtraction to define the soft factor associated with the jet,
\begin{eqnarray}
S_J(b_\perp)=\frac{S_{n_1,\bar n}(b_\perp)}{\sqrt{S_{n,\bar n}(b_\perp)}}\ ,
\end{eqnarray}
where $n_1$ represents the jet direction. 
A one-loop calculation leads to the following result:
\begin{eqnarray}\label{eq:soft-ave}
S_J^{(1)}(b_\perp,\mu_F)&=&\frac{\alpha_s}{2\pi}C_F\left[-\ln\frac{\hat t}{\hat u R^2}
\ln\frac{\mu_F^2}{\mu_b^2}-
\frac{1}{2}\ln^2\frac{1}{R^2}
\right] , \ \ \
\end{eqnarray}
where $\mu_b^2=c_0^2/b_\perp^2$, $c_0=2{\mathrm{e}}^{-\gamma_E}$ with the Euler constant $\gamma_E$. In order to derive the above result, we have averaged over the azimuthal angle of the jet. This average does not factorize and our results will only be valid up to next-to-leading logarithmic (NLL) order $\alpha_s^n \ln^n b_\perp$~\cite{future}, which is the accuracy we achieve in our current work. To see this, we note that the dependence on the azimuthal angle of the unaveraged soft function at NLO is non-logarithmic and thus enters the Wilson coefficient only in the full evolution formula. When integrating over the jet azimuthal angle, the Wilson coefficient reduces to its averaged form. Therefore, at NLL this is equivalent to performing the resummation using Eq.~(\ref{eq:soft-ave}) directly.

From the above result, we obtain the anomalous dimension $\gamma_s^{(1)}=-\ln(\hat t/\hat u R^2)C_F\alpha_s/2\pi$. Together with the result of the quark distribution from Ref.~\cite{Collins:2011zzd,Sun:2013hua}, the TMD factorization is verified at the one-loop order~\cite{future}. For anti-k$_T$ jets~\cite{Cacciari:2008gp}, the hard factor is given by
\begin{eqnarray}\label{eq:hard}
H_{\mathrm{TMD}}^{(1)}(Q,k_{\ell\perp})&=&\frac{\alpha_s}{2\pi} C_F\left[-\ln^2\frac{k_{\ell\perp}^2}{Q^2}-8-3\ln\frac{k_{\ell\perp}^2}{Q^2}\right.\nonumber\\
&&\left.\hspace*{-1cm}+\frac{3}{2}\ln\frac{1}{R^2} +
\frac{1}{2}\ln^2\frac{1}{R^2}
+\frac{13}{2}-\frac{2}{3}\pi^2\right] \,,
\end{eqnarray}
where we have chosen $\zeta_c^2=\hat s$ and $\mu_F^2=k_{\ell\perp}^2$ to simplify the expression. 

The large logarithms in the TMD quark distribution and the soft factor can be resummed by solving the relevant evolution equations. The result can be written as
\begin{eqnarray}\label{vq}
\widetilde{W}_q
|^{\mathrm{(resum.)}}&=&x\tilde{f}_{q}(x,b_\perp,\zeta_c=\sqrt{\hat s},\mu_F=k_{\ell\perp}) {\mathrm{e}}^{\Gamma_s(b_\perp)}H,\;\;\;\;\;\;\;\label{e11}
\end{eqnarray}
where the TMD quark distribution $\tilde f_q$ now contains the all order resummation and $\Gamma_s$ is the corresponding exponent for the soft factor,
\begin{equation}
\Gamma_s(b_\perp)=\int_{\mu_b^2}^{k_{\ell\perp}^2}
\frac{d\mu^2}{\mu^2}\gamma_s\big(\alpha_s(\mu)\big) \, ,
\end{equation}
with the one-loop result of $\gamma_s$ obtained above. In Eq.~(\ref{e11}), the hard factor $H$ contains finite terms from both the soft and hard factors in the TMD factorization in Eqs.~(\ref{eq:soft-ave}) and~(\ref{eq:hard}). Note that in particular the double logarithmic terms $\sim\ln^2(1/R^2)$ cancel out. We note that it is possible to further factorize Eq.~(\ref{eq:TMDfactorization}) in order to jointly resum logarithms of the jet radius $R$ following the techniques developed in~\cite{Kang:2016mcy,Liu:2017pbb}. However, at the EIC a large jet radius $R\sim{\cal O}(1)$ will be advantageous; see also~\cite{Gutierrez-Reyes:2018qez}.

Starting at two loops, non-global logarithms (NGLs) start to contribute to the cross section~\cite{Dasgupta:2001sh,Dasgupta:2002bw}. The leading contribution at order ${\cal O}(\alpha_s^2)$ is~\cite{Banfi:2003jj,future}
\begin{equation}
    S_{\mathrm{NGL}}^{(2)}(b_\perp)=-C_F\frac{C_A}{2}\left(\frac{\alpha_s}{\pi}\right)^2\frac{\pi^2}{24}\ln^2 \left(\frac{k_{\ell\perp}^2b_\perp^2}{c_0^2}\right) \ .
\end{equation}
The resummation of these NGLs is more complicated than that of the global logarithms captured in the resummation formula in Eq.~(\ref{vq}). For the numerical calculations presented below, we include their contribution by substituting in Eq.~(\ref{vq})
\begin{equation}
    \widetilde{W}_q\,\Longrightarrow\, \widetilde{W}_q \,{\cal S}_{\mathrm{NGL}}(b_\perp) \ ,
\end{equation}
where we use a simple parametrization ${\cal S}_{\mathrm{NGL}}(b_\perp)$ for the NGL contribution at leading color~\cite{Dasgupta:2001sh,Dasgupta:2002bw}. 

{\it 3. Single Transverse Spin Asymmetries as a Probes of the Quark Sivers Effect.}
Experimentally, the distribution of the total transverse momentum $q_\perp$ can be studied through the azimuthal angular correlation between the final state lepton and jet, and the uncertainty is better controlled than for $q_\perp$ itself. 
\begin{figure}[t]
\begin{center}
\includegraphics[width=7cm]{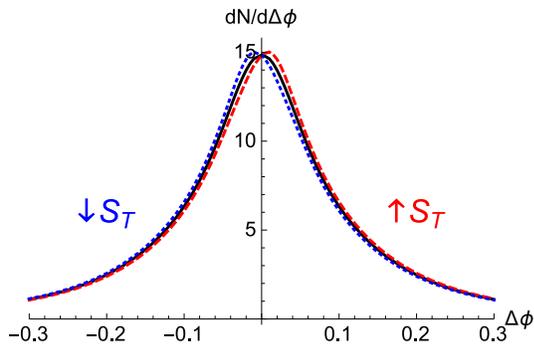}
\end{center}
\caption[*]{The azimuthal angular correlation between the final state lepton and jet at equal rapidities $y_\ell=y_J=1$ in the CM frame of the $ep$ collisions at $\sqrt{S_{ep}}=80$~GeV. We show $\Delta\phi=\phi_J-\phi_\ell-\pi$, where $\phi_J$ and $\phi_\ell$ are the azimuthal angles of the jet and lepton, respectively. We choose $k_{\ell\perp}=15$~GeV and integrate the jet transverse momentum over $10-20$~GeV, with radius $R=0.5$. The red and blue curves show the correlations when the spin of the transversely polarized nucleon is parallel or anti-parallel to $\vec{k}_{\ell\perp}$, respectively.}
\label{base}
\end{figure}
In Fig.~\ref{base}, we show this correlation for typical kinematics at the EIC with $\sqrt{S_{ep}}=80$~GeV, $k_{\ell\perp}=15$~GeV, $P_{J\perp}=10-20$~GeV, and we choose equal rapidities, $y_{\ell}=y_J=1$. The TMD quark distribution takes the form~\cite{Prokudin:2015ysa}, 
\begin{eqnarray}
\widetilde{f}_{q}(x,b_\perp,\zeta_c=\sqrt{\hat s},\mu_F=k_{\ell\perp})= {\mathrm{e}}^{-{ {S}_{\rm pert}^q(b_*)}-{S}_{\rm NP}^q(b_\perp)}\nonumber\\
\times\sum_iC_{q/i}(x,\mu_b/\mu)\otimes f_i(x,\mu_b)\ , \label{tmdqfc}
\end{eqnarray}
where $b_*=b_\perp/\sqrt{1+b_\perp^2/b_{\rm max}^2}$ with $b_{\rm max}=1.5$~GeV$^{-1}$, and $f_i(x,\mu)$ is the integrated parton distribution. Since there is only one TMD, the Sudakov form factor is given by
\begin{equation}
S_{\rm pert}^q(b_\perp)=\frac12 \int_{\mu_b^2}^{k_{\ell\perp}^2}
\frac{d\mu^2}{\mu^2}\left[A_q\big(\alpha_s(\mu)\big)\ln\frac{\hat s}{\mu^2}+B_q\big(\alpha_s(\mu)\big)\right] \ ,
\end{equation}
with $A_q=\frac{\alpha_s}{\pi}C_F$, $B_q=-\frac{\alpha_s}{\pi}\,\frac{3}{2}C_F$, and where for simplicity we take the leading order expression for the coefficient function $C$. We use the non-perturbative parametrization of Refs.~\cite{Su:2014wpa,Prokudin:2015ysa}: $S_{\rm NP}^q=0.106\, b_\perp^2+0.42\ln(Q/Q_0)\ln(b_\perp/b_*)$ with $Q_0^2=2.4$~GeV$^2$. The result in Fig.~\ref{base} is shown as a function of $\Delta\phi=\phi_J-\phi_\ell-\pi$ which can be viewed as a measure of the decorrelation away from the back-to-back configuration. As expected, the distribution peaks around $\Delta\phi=0$ where the broadening effects depend on the TMDs. We also find that the NGL contribution is very small for most of the kinematic range considered here except for $\Delta\phi\sim 0$, where it yields a suppression of about $5\%$.

When the nucleon is transversely polarized, the TMD quark distribution will have an azimuthal asymmetry due to the Sivers effects. As a result, the azimuthal angular distribution will no longer be symmetric with respect to $\Delta \phi=0$~\cite{Boer:2003tx}. The deviation probes the size of the quark Sivers function. The transverse-spin dependent differential cross section can be written as
\begin{eqnarray}
\frac{d\Delta\sigma(S_\perp)}{dy_\ell d^2 k_{\ell \perp} d^2q_{\perp}}=\sigma_0\epsilon^{\alpha\beta}S_\perp^\alpha\int\frac{d^2b_\perp}{(2\pi)^2}{\mathrm{e}}^{iq_\perp\cdot b_\perp} \widetilde{W}_{Tq}^\beta\ .
\end{eqnarray}
The spin-dependent $\widetilde{W}_{Tq}^\beta$ can be factorized into the Sivers quark distribution $\tilde{f}_{1T}^{\perp\beta}$ and the soft and hard factors as
\begin{eqnarray}
    \widetilde{W}_{Tq}^\beta
    =x\tilde{f}_{1T}^{\perp\beta}(x,b_\perp)\,
    S_J(b_\perp,\mu_F)\, H_{\mathrm{TMD}}^\perp(Q,\mu_F)\ .
\end{eqnarray}
Again, a one-loop calculation can be carried out for this observable. Most of the calculation is similar to that for SIDIS~\cite{Ji:2006br,Koike:2007dg}, except for the dynamics associated with the observed jet. The hard factor turns out to be the same as in the unpolarized case, and we leave a detailed derivation for future work~\cite{future}. 

For the numerical calculations, we use the parametrization of the quark Sivers function of Ref.~\cite{Sun:2013hua},
\begin{eqnarray}
    \tilde{f}_{1T}^{\perp\beta}(x,b_\perp,\zeta_c,\mu_F)&=&\frac{ib_\perp^\beta M}{2}N_q\frac{x^{\alpha_q} (1-x)^{\beta_q}}{\alpha_{q}^{\alpha_q}\beta_q^{\beta_q}}\nonumber\\
    &&\times \tilde f_q(x,b_\perp,\zeta_c,\mu_F){\mathrm{e}}^{g_sb_\perp^2}\ ,
\end{eqnarray}
where $g_s=0.062$~GeV$^2$. Here we assume that the twist-three correlation of the quark Sivers function (transverse momentum moment) has a similar scale dependence as the integrated quark distribution, $f_q(x,\mu_b)$ in Eq.~(\ref{tmdqfc}), which will introduce a small theoretical uncertainty 
in the kinematics of interest at low transverse momentum~\cite{Sun:2013hua}. This could be improved in the future by taking into account the complete scale evolution of the twist-three functions~\cite{Braun:2009mi,Kang:2008ey,Vogelsang:2009pj,Zhou:2008mz,Schafer:2012ra}.

\begin{figure}[t]
\begin{center}
\includegraphics[width=7cm]{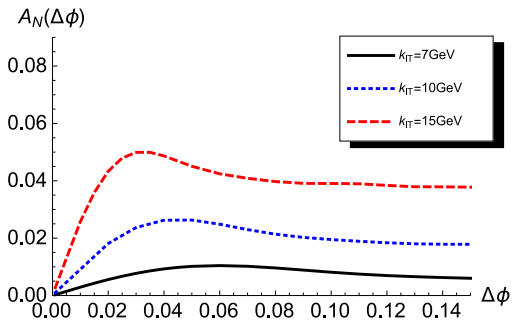}
\end{center}
\caption[*]{The single transverse spin asymmetry as function of $\Delta \phi=\phi_J-\phi_\ell-\pi$ for different lepton transverse momenta $k_{\ell\perp}=7$, 10, and 15~GeV, respectively, which illustrates the transverse momentum dependence of the quark Sivers function.}
\label{ssa}
\end{figure}

As an illustration of the Sivers effect we show the modifications of the correlation spectrum in Fig.~\ref{base} when the transverse spin of the nucleon $\vec{S}_\perp$ is parallel or anti-parallel to the transverse momentum $\vec{k}_{\ell\perp}$ of the final state lepton. In  Fig.~\ref{ssa}, we show the single spin asymmetry directly as a function of $\Delta\phi$. The $\Delta\phi$ distribution provides information on the transverse momentum dependence of the Sivers function as $\Delta \phi$ is proportional to $q_\perp$ in the correlation limit. The magnitude of the asymmetry depends on how large the Sivers function is compared to the unpolarized quark distribution. For example, for $k_{\ell\perp}=7$~GeV the asymmetry is very small because the probed value of $x$ is about 0.03 and the quark Sivers function is very small. Therefore, by varying the lepton's momentum and the rapidities of the lepton and jet, we can study the $x$-dependence of the asymmetry, which will lead to significant constraints on the quark Sivers function. Compared to the Sivers asymmetry in SIDIS, this observable has the advantage that it does not involve TMD fragmentation functions. The asymmetry itself can directly provide information the size of the quark Sivers function relative to its unpolarized counterpart. This will provide a unique opportunity for the transverse momentum tomography of the quark in a transversely polarized nucleon at the EIC. From the size of the effect we expect that it will observable at the EIC and allow
detailed scans of the $k_\perp$ and $x$ dependences of the Sivers functions with a clear physics interpretation within TMD factorization.

Extending the above calculation to the dijet spin asymmetry in $pp$ collisions~\cite{Boer:2003tx} will be very interesting. In particular, the previous calculations~\cite{Vogelsang:2005cs,Bomhof:2007su} should be improved by taking into account QCD evolution effects. Using the correlation of dijets, it will be possible to explore factorization breaking effects and the non-universality of the Sivers functions by comparing to the available experimental data~\cite{Abelev:2007ii}.

{\it 4. $P_T$-Broadening  as a Probe of Cold Nuclear Matter Effects.} 
As a second example we will show that the process in (\ref{eq:process}) can be used to explore cold nuclear effects in $eA$ collisions at the EIC. When a highly energetic jet is produced in the hard partonic process, it experiences multiple interactions with the target nucleus which will generate $P_T$-broadening effects~\cite{Baier:1996sk}. These final state interactions can also be factorized into the TMD quark distribution of the nucleus~\cite{Liang:2008vz}. 

As shown in Refs.~\cite{Mueller:2016gko,Mueller:2016xoc}, nuclear $P_T$-broadening effects can be systematically included within the framework of TMD resummation by modifying $\widetilde{W}_q$  
as
\begin{equation}
\widetilde{W}_q\,\Longrightarrow\, \widetilde{W}_q\, {\mathrm{e}}^{-\frac{\hat qLb_\perp^2}{4}} \ ,
\end{equation}
where $\hat qL$ represents the typical transverse momentum obtained by the quark through multiple interactions with the cold nuclear matter. On the right hand side of the above equation, the first factor contains the intrinsic contribution from the nucleon and the Sudakov exponent associated with QCD evolution. We can combine the nucleon's intrinsic contribution with the $\hat qL$-term to represent the TMD contribution from the nucleus~\cite{Liang:2008vz}. This shows that we have a consistent picture of $P_T$-broadening effects for the process in (\ref{eq:process}) in $eA$ collisions. 

\begin{figure}[t]
\begin{center}
\includegraphics[width=7cm]{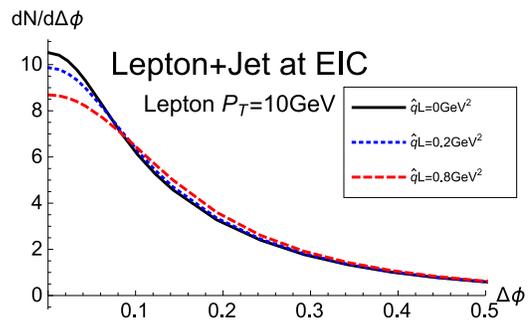}
\end{center}
\caption[*]{$P_T$-broadening effects for the lepton jet azimuthal correlation due to the interaction with cold nuclear matter as a function of $\Delta \phi=|\phi_J-\phi_\ell-\pi|$ for two typical values of 
$\hat q L$.}
\label{ea}
\end{figure}

In Fig.~\ref{ea}, we plot the azimuthal angular correlation as a function of $\Delta \phi$ for different values of $\hat qL$. These values are in the range of a theoretical estimate for cold nuclear matter~\cite{Baier:1996sk}. We expect that this correlation can be investigated at the future EIC, and the comparison with the dijet correlation measurement in heavy-ion collisions~\cite{Chen:2016vem,Chen:2016jfu,Chen:2018fqu,Tannenbaum:2017afg,Adamczyk:2017yhe,Aad:2010bu,Chatrchyan:2011sx} will provide important information on hot and cold dense nuclear matter using hard probes.

{\it 5. Summary.}
We have proposed lepton-jet correlations as a new class of observables in DIS at the future electron-ion collider for nucleon/nucleus tomography. In particular, we have demonstrated that the single transverse spin asymmetries for this process directly probe the quark Sivers function. In $eA$ collisions, the measurement of $P_T$-broadening effects will be a great opportunity to explore cold nuclear matter effects through hard probes.

The advantage of the lepton-jet correlation as compared to the standard SIDIS processes is that it does not involve TMD fragmentation functions. Extensions to other observables that are sensitive to the various TMD quark distributions at leading order shall follow. For $eA$ collisions, we can also study jet energy loss in the process (\ref{eq:process}). The comparison with heavy ion data~\cite{Adamczyk:2017yhe,Aad:2010bu,Chatrchyan:2011sx} will shed light on the underlying physics mechanisms for jet energy loss in hot and cold QCD matter. We hope that the results presented in this work will stimulate further theoretical developments along these directions.

{\it Acknowledgement.}
We thank Iris Abt, Elke-Caroline Aschenauer, Brian Page and Ernst Sichtermann for useful discussions on the experimental perspectives of the process studied in this paper. F.Y. thanks Yoshitaka Hatta for discussions on the non-global logarithmic contributions. The material of this paper is based upon work partially supported by the LDRD program of Lawrence Berkeley National Laboratory, the U.S. Department of Energy, Office of Science, Office of Nuclear Physics, under contract number DE-AC02-05CH11231.

\end{document}